\documentclass[prb,preprint]{revtex4-1} 

\usepackage{amsmath}  
\usepackage{amsfonts} 
\usepackage{graphicx} 

\begin{document}


\title{Quadrupole Excitations in hysteresis loops of magnetic NanoCluster $Fe_8$}

\author{Yousef Yousefi}
\email{Yousof54@yahoo.com} 
\altaffiliation[permanent address: ]{Payame Noor University
 POB 19395-4697,Tehran,Iran.  
  } 
\affiliation{Department of Physics, Payame Noor University, Tehran, IRAN}

\author{H. Fakhari}
\email{Yousof54@gmail.com}
\affiliation{Department of Physics, Payame Noor University, Tehran, IRAN}


\date{\today}

\begin{abstract}
Experiments show several steps in hysteresis loops of high spin nanocluster $Fe_8$. It is thought that these steps are due to thermally assisted resonant tunneling between different quanta spin states. Up to now, in calculatiing it, only dipole excitations were considered. Because of the symmetry and the power of spin operators in Hamiltonian, we think that other multipole excitation must be considered too. In this paper we consider both dipole and quadrupole excitations in Hamiltonian and then quantum resonant tunneling phenomena are obtained numerically. As we can show in these figures, this phenomenon is different in these two conditions and the second condition is nearly to the fact.
\end{abstract}

\maketitle 

\section{Introduction} 

Today more than five thousand different magnetic molecular clusters are known, and about hundred of these show some behaviors that are significant from physics science point of view [1]. These molecules are between simple paramagnetic salts and super-paramagnetic particles of sub-micrometer size. Thus, they show phenomena such as hysteresis at molecular level. Some issues in this subject are similar to those that result in the study of magnetization reversal of small magnetic particles, and it is hoped that the molecular systems will offer insights into the future, which has applications in magnetic recording and storage technologies.

The molecule  $Fe_8$ (proper chemical formula: $[Fe_8 O_2 (OH)_{12} (tacn)_6]^{8+}$ is magnetic, and forms good single crystals.(see Fig.1). In its lowest state, it has a total spin of 10, arising from competing antiferromagnetic interactions between the eight $S = 5/2$ Fe ions within a molecule. Spin-orbit and spin-spin interactions destroy complete rotational invariance, and give rise to anisotropy with respect to the crystal lattice directions. A variety of experimental techniques (electron spin resonance, ac susceptibility, magnetic relaxation, Mossbauer spectroscopy, neutron scattering) indicates that their data can be  fitted by the following model Hamiltonian $[2 - 5]$:

\begin{equation}
\mathcal{H} =-DS^2_z -ES^2_x-g \mu_BS.H
\end{equation}

Where $S_x$ and $S_z$ are the two components of spin operator, D and E are the anisotropy constants and are known through a variety of experimental evidences with$ D=0.33k$,  $E=0.092k$ [6-7]. The last term of the Hamiltonian describes the Zeman energy which is associated with an applied field H. This Hamiltonian defines hard, and easy axes of magnetization in x and z direction respectively. 

 Magnetization hysteresis are exhibited for  this nanocluster below blocking temperature . Fig.2 shows typical hysteresis loops with the field applied along the easy axis of magnetization. The steps in the loops are clearly visible. These loops become temperature independent below 0.35 k and demonstrate quantum tunneling at the lowest levels of energy.

\begin{figure}[htb]
\begin{minipage}{14pc}
\includegraphics[width=1\textwidth]{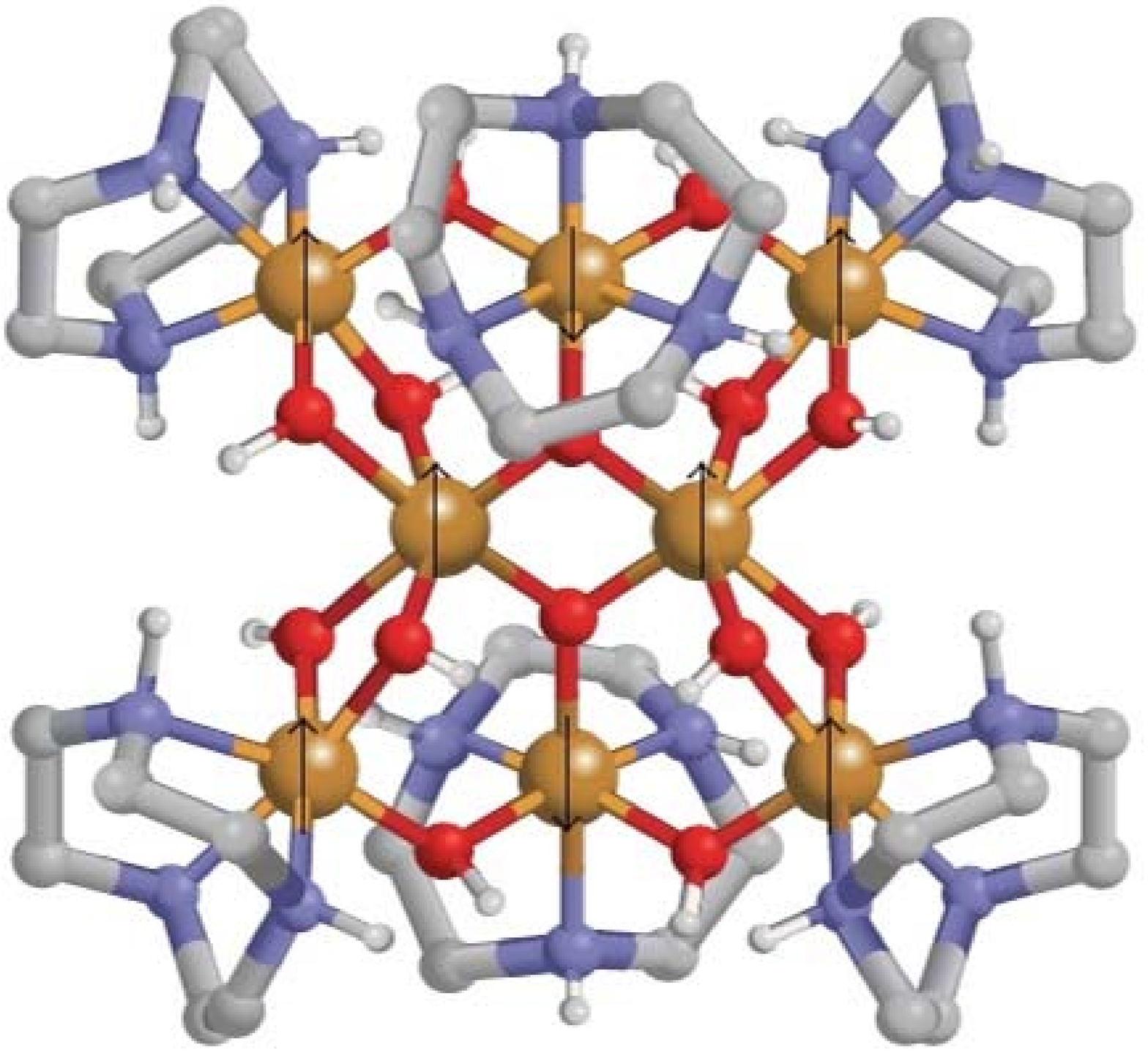}
\caption{\label{label} Chemical structure  in nanocluster $Fe_8$.}
\end{minipage}\hspace{2pc}%
\begin{minipage}{14pc}
\includegraphics[width=1\textwidth]{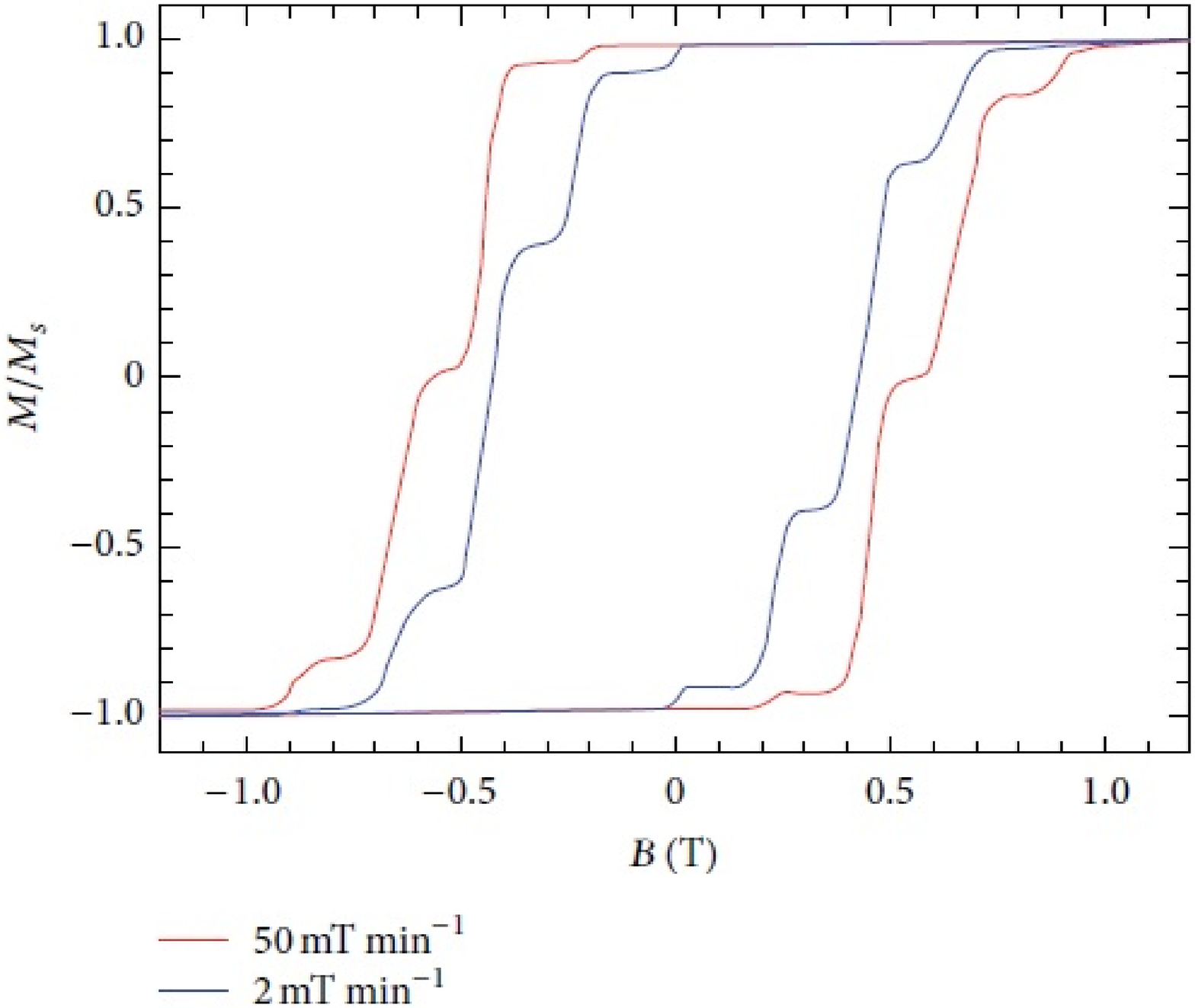}
\caption{\label{label}hysteresis loops with the field applied along the easy axis of magnetization.}
\end{minipage} 
\end{figure}

It suggested that the observed steps in hysteresis loops are due to resonance tunneling between quantum spin states. In calculating only dipole excitations were considered [8] and we think that other  the excitations, such as  multipole excitations are related, and must be considered.

In this paper, firstly, periodically dependence of spin tunneling for $Fe_8$ in SU(2) group is calculated. In other word, only dipole excitation in Hamiltonian is considered. Due to the symmetry and the power of spin operators in Hamiltonian and  being more accurate, other multipole excitations must be considered. Then, in this study we consider both dipole and quadrupole excitations. Instanton method is used to calculate tunneling phenomenon between the ground states. This method will be described in the next section.

\section{Theory}

Instantons are classical paths that run between degenerating the classical minima of the energy. By classical we mean that the path obeys the principle of the  least action and satisfies energy conservation. However, a path along which energy is conserved cannot run between two minima and can have real coordinates and momenta. Hence, one must enlarge the notion of a classical path and allow the coordinates and/or momenta to become complex.

The instanton method is an efficient way of calculating tunnel splitting, both for particles [9] and for spin [10]. It is based on evaluating the path integral for a certain propagator in the steepest-descent approximation and is designed to be asymptotically correct in the semiclassical limit $J \rightarrow \infty$ or $ \bar h \rightarrow 0$.

The tunneling amplitude in this method is given by [11]:

\begin{equation}
\triangle =\sum _k D_k e^{-S^{cl}_k}
\end{equation}

The pre-factor $D_k$  results from integrating the Gaussian fluctuations around the k-th instanton.
For nanocluster $Fe_8$, the tunneling amplitude is obtained from the following relation [5]:

\begin{equation}
\triangle \approx 2 D_1 e^{-\mathcal{A}}cos(J\Theta) 
\end{equation}

Where $\mathcal{A}$ and $\Theta$ related to real and imaginary parts of classical action. In this paper, these functions were  calculated numerically. Because this pre-factor don't effect on the location of quenching points or steps in hysteresis loops, for simplicity we assume $2D_1=1$.

\section{Calculations in SU(2) group}

We used expectation value of operators $S_i$ in SU(2) group to obtain classical energy of the system, in other word, in the first case only dipole excitations are considered. Then

\begin{eqnarray}
S^+ &=& e^{i\phi} sin\theta\nonumber\\                                                                                                     
S^- &=& e^{-i\phi} sin\theta \nonumber\\                                                                                                
S^z &=& cos\theta                                                                                                                                     
\end{eqnarray}

By placing these values in Hamiltonian (1.1) obtained:

\begin{equation}
E_{cl} =-D cos^2\theta +Esin^2\theta sin^2\phi-A cos\theta 
\end{equation}

Where $A =10g\mu_B H$. The minimum of energy is obtained at:

\begin{equation}
\theta=\pi/2 , {   }   \phi=0,\pi
\end{equation}

The minima of energy are:

\begin{eqnarray}
E_{cl}(\pi/2,0)&=& 0\nonumber\\                                                                                                     
E_{cl}(\pi/2,\pi) &=& 0                                                                                
\end{eqnarray}

By substituting $P=cos\theta$ , the instanton must satisfy the constraint:

\begin{equation}
-D P^2-AP +E(1-P^2)sin^2\phi=0 
\end{equation}

Solutions for the above equation are:

\begin{equation}
P(\phi)=\frac{-A \pm\sqrt{A^2+4E sin^2 \phi (D+E sin^2\phi)}}{2(D+E sin^2\phi)}
\end{equation}

The above solutions, despite conserving energy, must satisfy the boundary conditions $ P(0)=P(\pi)=0$ . These solutions give rise to what A. Garg called boundary jump instantons [3]. The solution whcih satisfies the above conditions is:

\begin{equation}
P(\phi)=\frac{-A +\sqrt{A^2+4E sin^2 \phi (D+E sin^2\phi)}}{2(D+E sin^2\phi)}
\end{equation}
                                                                       
If we consider only dipole excitations, $\mathcal{A}$ and $\Theta$ obtain in the following forms:

\begin{eqnarray}
\mathcal{A}&=&ReS_{cl}=Re \int_0^{\pi} P(\phi)d\phi \nonumber\\
\Theta &=& \int_0^{\pi} (1-P(\phi))d\phi
\end{eqnarray}

We calculated these functions numerically  and substituted in relation (2.2) and tunneling amplitude  obtained in the form of Fig. 4. If we plotted function $cos(J\Theta)$ versus magnetic field h, the Fig. 3 would be obtain.

\begin{figure}[htb]
\begin{minipage}{14pc}
\includegraphics[width=1\textwidth]{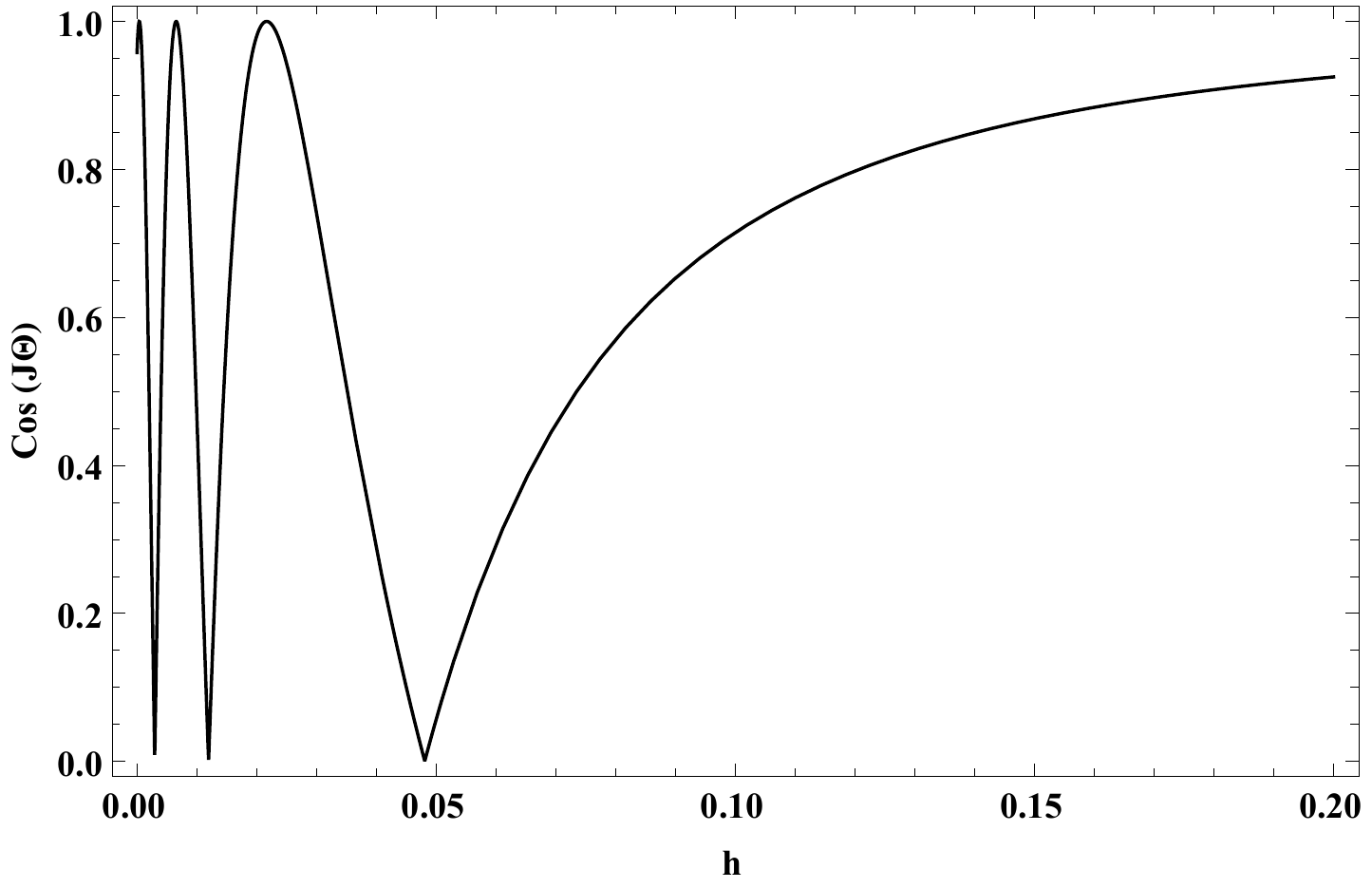}
\caption{\label{label} $cos(J\Theta)$ versus magnetic field h, in this case only dipole excitations is considered.}
\end{minipage}\hspace{2pc}%
\begin{minipage}{14pc}
\includegraphics[width=1\textwidth]{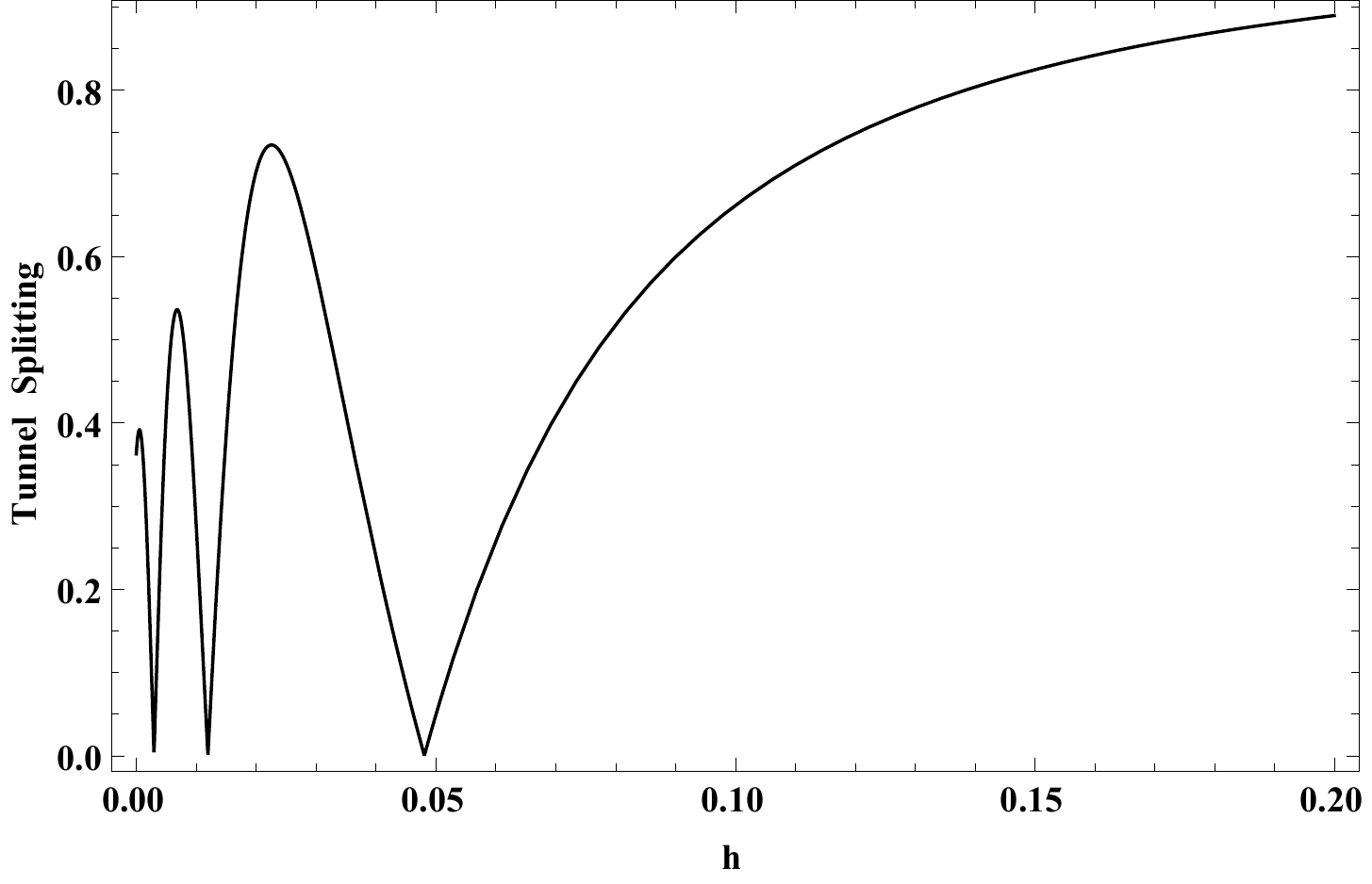}
\caption{\label{label}Amplitude of tunneling versus magnetic field h, in this case only dipole excitations is considered.}
\end{minipage} 
\end{figure}

This graphic show that there are three quenching points in spin tunneling amplitude, but we know from experimented date that the number of these points or steps in hysteresis loop are four and for this reason other multipole excitations must be considered. In the next section, we will add quadrupole excitation to calculations and obtain these functions and graphics.

\section{Calculations in SU(3) group}

If quadrupole excitation is considered, we use spin expectation values in SU(3) group:

\begin{eqnarray}
S^+ &=& e^{i\phi}cos2g sin\theta\nonumber\\                                                                                                     
S^- &=& e^{-i\phi}cos2g sin\theta \nonumber\\                                                                                                
S^z &=& cos2gcos\theta                                                                                                                                     
\end{eqnarray}

If placing these quantities in Hamiltonian (1.1), the classical energy would obtain in the following form:

\begin{equation}
E_{cl} =-D cos^2 2gcos^2\theta +Ecos^2 2gsin^2\theta sin^2\phi-Acos 2g cos\theta 
\end{equation}
The minimum of energy obtained at:

\begin{equation}
\theta=\pi/2 , g=\pi/4 ,  \phi=0,\pi
\end{equation}
Also minima of energy are:
\begin{eqnarray}
E_{cl}(\pi/2, \pi/4,0)&=& 0\nonumber\\                                                                                                     
E_{cl}(\pi/2, 5\pi/4,\pi) &=& 0                                                                                
\end{eqnarray}

By substituting  $P=cos\theta$ , the instanton must satisfy the constraint:

\begin{equation}
-D cos^2 2g P^2 +Ecos^2 2g(1-P^2)sin^2\phi-Acos2gP=0 
\end{equation}
Solutions for the above equation are:
\begin{equation}
P(\phi, g)=\frac{-A \pm\sqrt{A^2+4Ecos^2 2g sin^2 \phi (D+E sin^2\phi)}}{2cos2g(D+E sin^2\phi)}
\end{equation}
Similar to the previous section, solution that has the following form is considered:

\begin{equation}
P(\phi, g)=\frac{-A +\sqrt{A^2+4Ecos^2 2g sin^2 \phi (D+E sin^2\phi)}}{2cos2g(D+E sin^2\phi)}
\end{equation}

\begin{figure}[htb]
\begin{minipage}{14pc}
\includegraphics[width=1\textwidth]{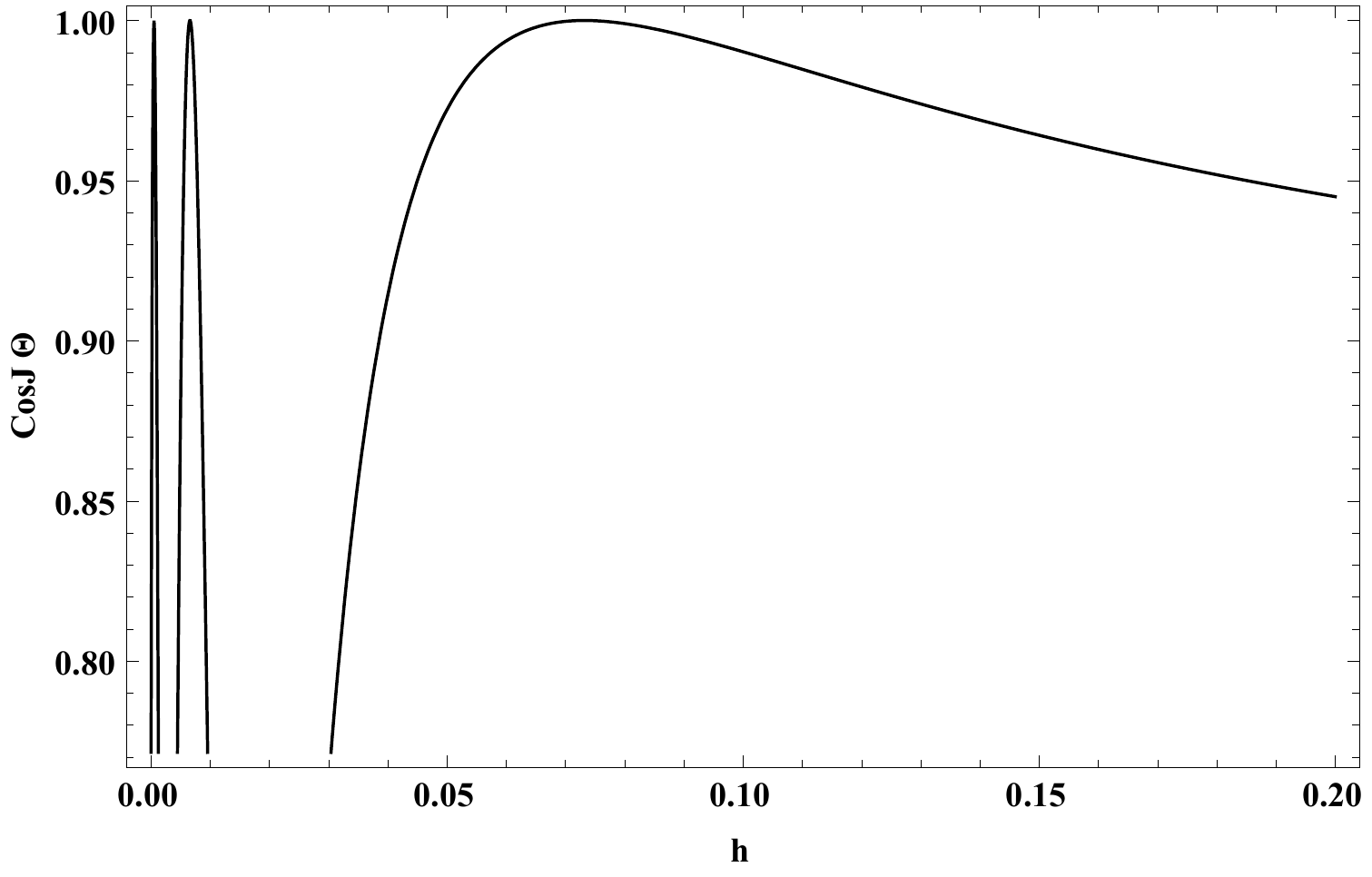}
\caption{\label{label} $cos(J\Theta)$ versus magnetic field h, in this case both dipole and quadrupole excitations  are considered.}
\end{minipage}\hspace{2pc}%
\begin{minipage}{14pc}
\includegraphics[width=1\textwidth]{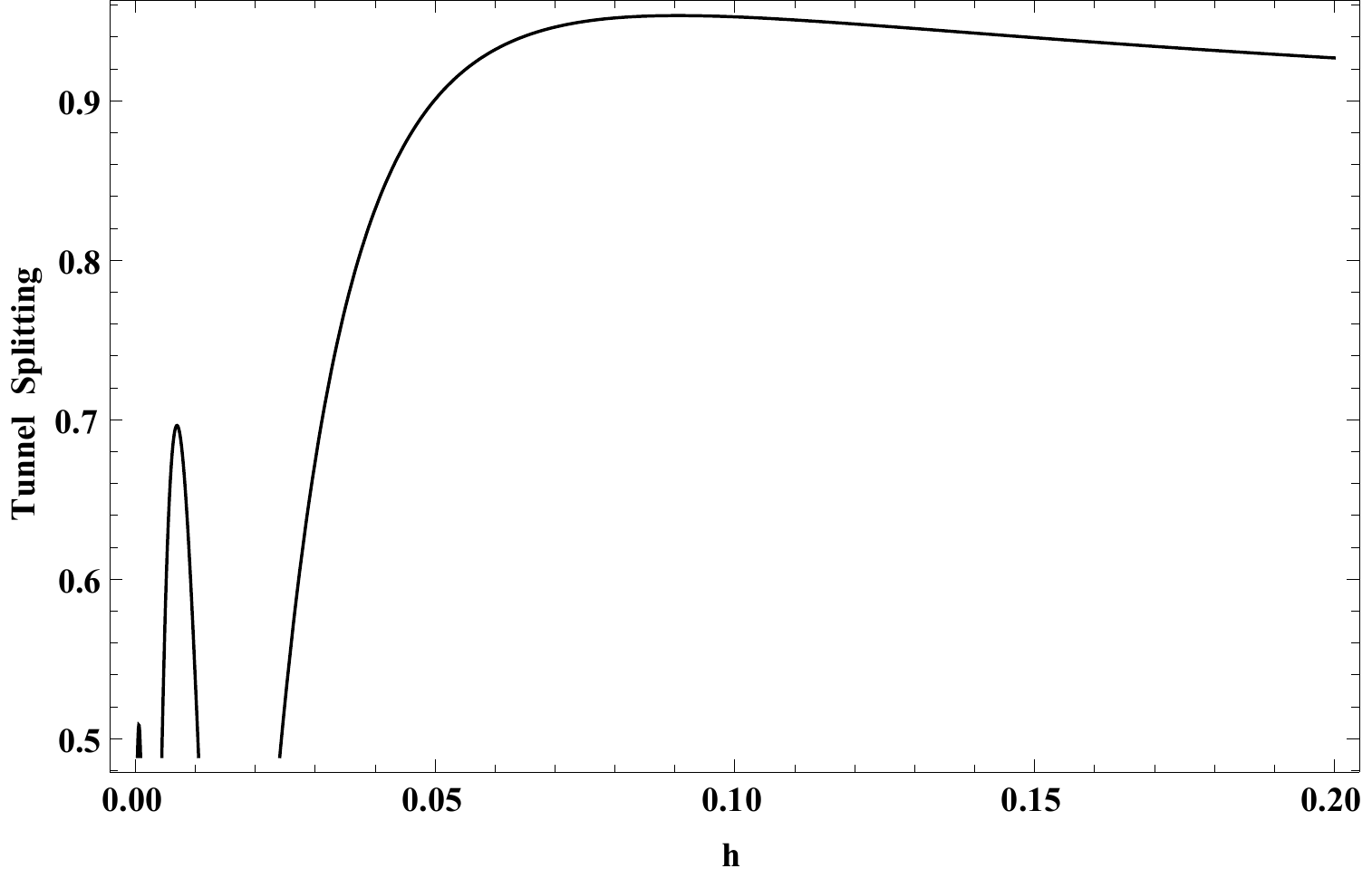}
\caption{\label{label}Amplitude of tunneling versus magnetic field h,in this case both dipole and quadrupole excitations  are considered.}
\end{minipage} 
\end{figure}

If we consider both dipole and quadrupole excitations, $\mathcal{A}$ and $\Theta$  will obtain in the following forms:

\begin{eqnarray}
\mathcal{A}&=&ReS_{cl}=Re \int_0^{\pi} \int_0^{\pi/4} P(\phi,g)d\phi dg \nonumber\\
\Theta &=& \int_0^{\pi} (cos2g-P(\phi,g))d\phi- \int_0^{\pi} (1-cos2g)d\gamma
\end{eqnarray}

Where two variables $\theta$ and $\phi$ are related to dipole excitations and g and $\gamma$ are related to quadrupole excitations.

We calculated  this function numerically and substituted them in relation (2.2). If we plotted tunnel amplitude and  function $cos(J\Theta)$ versus magnetic field h, quenching points in splitting (Fig. 5) or steps in hysteresis loop (Fig. 6) would be obtained.

As we saw in this plot, the number of quenching points  for this Nanocluster are equal to the number of steps in hysteresis loop.

\section{Discussion}

In this paper, according to the power of spin operators in Hamiltonian,  to get the correct number of steps in hysterics loop, we used  dipole and quadrupole excitations.

At the beginning, only dipole excitation was considered, as we saw in Fig. 4, the numbers of steps in hysterics loop or quenching points in tunnel splitting oscillations are three and this number is less than real one. If we add quadrupole excitation, this number changes to four, Fig. 6, and this is equal to the fact.

Note that, it is possible that the position of these steps or quenching points is different from the real state. To be more accurate, we must add another terms as perturbed terms to Hamiltonian.

\end{document}